\documentclass[a4paper,12pt]{article}
\usepackage[stable]{footmisc}
\usepackage{amsmath,graphicx}     
\def\twiddles#1{\mathrel{\mathop{\sim}\limits_
                        {\scriptscriptstyle {#1\rightarrow 0 }}}}

\begin{document}

\begin{flushright}
  TIF-UNIMI-2020-1
\end{flushright}

\begin{center}{\Large On the Sudakov form factor,\\ and a factor of two}\\
\bigskip

\vspace{0.4 truecm}

{\bf Stefano Forte}
 \\[5mm]

{\it Tif Lab, Dipartimento di Fisica, Universit\`a di Milano and\\ INFN, Sezione di Milano,\\ Via Celoria 16, I-20133 Milano, Italy\\[2mm]}

\vspace{0.8cm}

{\bf Abstract }
\end{center}

I answer a question that Roman Jackiw asked me, and I draw some lessons from
the answer. The
question is: why is the Sudakov form factor larger by a factor of two,
if computed for off-shell fermions, in comparison to the on-shell
case?
The answer sheds some light on
the interplay between infrared and collinear singularities --- and the
importance of factors of two.

\begin{center}
Contribution to the volume {\it Roman Jackiw ---
 80th Birthday Festschrift}
\end{center}

\section{Master of scientific style}
Much can be said about working under Roman's supervision in the
mid-eighties: it  was an absorbing,
intense, at times exhilarating, at times stressful experience. A
formidable array of ideas to take in, concepts to grasp, and good
practices to learn,  often delivered as a side
remark, accompanied by a grin\footnote{See the contribution  by
  Michiel Bos in this volume.}.  Ranging from the way to write
displayed equations in a paper 
(``it is called an equation because it has an equal sign''), to the
importance of choosing the symbols when performing a calculation
(``one should not pick letters at random from the alphabet''). 
Some perks, too, such as going for dinner at the Harvard
faculty club with Steven Hawking and Sidney Coleman --- including the
task of steadying the former's wheelchair in the minivan that took us there.

Overall, it amounted to a lesson of scientific method, and of scientific style:
delivered mostly by example. One thing I understood --- the painful
way, as I am prone to algebraic mistakes --- is the importance of
details when performing a computation. Roman used to retell the story of
someone who published a perturbative computation in which he had
guessed the value of a high-order term without actually
calculating it --- only to be belied by the explicit result once Roman
got round to determine it.

Not so long ago, I came to think about it again.
It was the Summer of 2017,  I was spending some time at the Aspen center
for physics, and Roman, coincidentally also there, took me out for lunch. The conversation at some
point revolved on some then-recent work of mine~\cite{Muselli:2017bad}
on QCD resummation. Roman mentioned that he had
worked on related topics around the time of his PhD thesis~\cite{Jackiw:1968zz}: he had computed
perturbatively  the
high-momentum transfer limit of the QED vertex function, which is
double logarithmic, and correctly
guessed the exponentiation of the double logs. As our lunch was going
on,
Roman then
abruptly asked me whether I knew that the coefficient of the double
log is by a factor of two larger off-shell in comparison to the on
shell-result, and whether I knew a simple physical
reason for that.

I didn't know.

Continuing the discussion with Roman via email, after I got back home,
I realized that I couldn't immediately come up with
an answer. I also subsequently realized that this point is typically not
discussed, or even
mentioned in textbooks. In fact, what is commonly known as ``the
Sudakov form factor'' is the on-shell result, yet the
original\cite{Sudakov:1954sw} Sudakov calculation applies to the
off-shell case, and the factor two difference usually goes unnoticed. Indeed, as Roman pointed out to me, in a recent paper from the Russian
school~\cite{Ioffe:2012re} it is incorrectly stated\footnote{See in particular the discussion
  after Eq.~(34) of Ioffe's paper.~\cite{Ioffe:2012re}} that the same result
applies in the on-shell and off-shell cases, only with a different
choice of infrared regulator.  I asked various 
experts on QCD,
where the Sudakov exponentiation plays an important role, and none
was aware of this.

Answering Roman's question is the purpose of this note.

\section{The vertex function and\\ the Sudakov form factor}\label{sec:sudakov}
\begin{figure}[htb]
\centerline{\includegraphics[width=.35\linewidth]{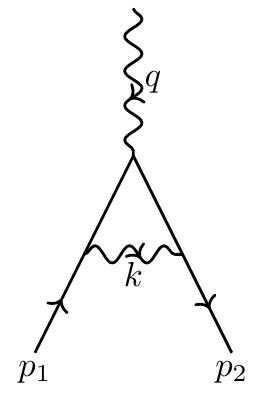}}
\caption{The one-loop vertex function.}
\label{fig:vertex}
\end{figure}

The computation performed by Roman\cite{Jackiw:1968zz}
determines to
all orders the high-energy 
behavior of the vertex function in Quantum Electrodynamics (QED)  (see Fig.~\ref{fig:vertex}). The genesis of this paper has been recounted
by Roman\cite{Jackiw:2013jba}: his advisor, Ken Wilson,
suggested him as a thesis project to derive this high-energy behavior, which had
been previously obtained by Sudakov\cite{Sudakov:1954sw} in the
off-shell case, using
renormalization-group (RG) methods:
both as a way of validating the then-novel RG
techniques, and also, of obtaining the on-shell result.

The way to attack and solve this problem using  RG
techniques was only found several years
later~\cite{Contopanagos:1996nh,Forte:2002ni} (see also Sect.~\ref{exp}
below), but Roman did manage to
tackle it by direct computation using an eikonal approximation, which
was known to Wilson, and systematically developed by
Weinberg~\cite{Weinberg:1966jm} into  what is now known as light-cone
field theory.\footnote{
  A textbook
  discussion of the eikonal approximation is e.g. given by
  Sterman\cite{Sterman:1994ce}, while a presentation of
  light-cone field theory can be found in recent summer school proceedings\cite{Venugopalan:1998zd}.}

The result found by Roman\cite{Jackiw:1968zz} is that in the high-energy
limit the vertex function  is equal to
\begin{equation}
  \Gamma^\mu(p_1,p_2)=\gamma^\mu
  \Gamma(p_1^2,p_2^2,q^2); \label{defgamma}
  \end{equation}
where to one loop  (Fig.~\ref{fig:vertex})
\begin{equation}
\Gamma^{(1),\, {\rm off}}
(p_1^2,p_2^2,q^2)=-\frac{\alpha}{2\pi}\ln\left|\frac{q^2}{p_1^2}\right|\ln\left|\frac{q^2}{p_2^2}\right| \label{offshellone}
\end{equation}
for off-shell fermions with virtualities $p_i^2$,
while for on-shell fermions
\begin{equation}
\Gamma^{(1),\, {\rm on}}
(m^2,m^2,q^2)=-\frac{\alpha}{4\pi}\ln^2\frac{\left|q\right|^2}{\mu^2} \label{onshellone}
\end{equation}
where $\mu$ is an infrared regulator (i.e. a photon mass). 

To all perturbative orders the one-loop result exponentiates:
\begin{align}
  \Gamma^{\rm off}
(p_1^2,p_2^2,q^2)&=\exp-\frac{\alpha}{2\pi}\ln\left|\frac{q^2}{p_1^2}\right|\ln\left|\frac{q^2}{p_2^2}\right|,\label{offshellall}
    \\
  \Gamma^{\rm on}
    (m^2,m^2,q^2)&=\exp-\frac{\alpha}{4\pi}\ln^2\frac{\left|q\right|^2}{\mu^2}.\label{onshellall}
\end{align}
All these results hold to double-logarithmic accuracy, i.e. up to
terms with a lower power of $\ln|q|^2$.

The result Eq.~(\ref{offshellall}) is in agreement with the previous result of
Sudakov\cite{Sudakov:1954sw}, which had been subsequently
reproduced by others~\cite{cassandro}, who attempted to determine the
on-shell result but did not obtain the correct answer and failed to
prove exponentiation. The results
Eqs.~(\ref{offshellall}-\ref{onshellall}) are also given in
the volume devoted to QED of ``Landau's'' theoretical physics
course\cite{Berestetsky:1982aq}, first published in 1974 (after
Landau's death): for the off-shell result Sudakov\cite{Sudakov:1954sw} is cited, while the
on-shell result is written in the form
\begin{equation}\label{onlandau}
\bar \Gamma^{(1),\,{\rm on}}
(m^2,m^2,q^2)=-\frac{\alpha}{4\pi}\left(\ln^2\left|\frac{q^2}{m^2}\right|+4\ln\left|\frac{q^2}{m^2}\right|\ln\left|\frac{m}{\mu}\right|\right),
\end{equation}
which of course coincides with Eq.~(\ref{onshellone}) up to terms
which are not logarithmic in $q^2$:
\begin{equation}
\bar \Gamma^{(1),\,{\rm on}}(m^2,m^2,q^2)= \Gamma^{(1),\,{\rm
    on}}(m^2,m^2,q^2) +\frac{\alpha}{4\pi}\ln^2 \frac{m^2}{\mu^2}\label{jacklan}.
\end{equation}

It is clear that, contrary to what one might naively think (and contrary
to what sometimes  stated explicitly\cite{Ioffe:2012re}) the on-shell result is not simply
 obtained by setting $p_1^2=p_2^2=\mu^2$ in the off-shell one --- rather, it
 is twice as large. Why?

\begin{figure}[htb]
\centerline{\includegraphics[width=.6\linewidth]{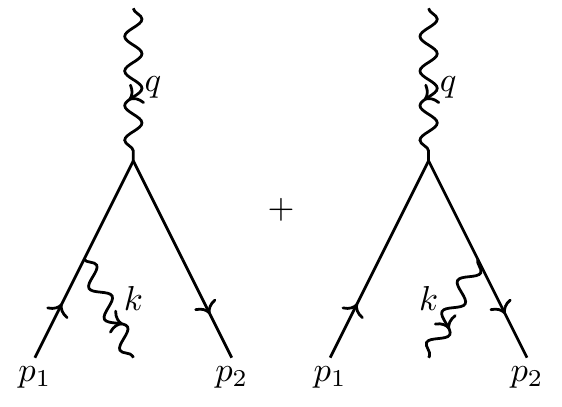}}
\caption{The 
  real emission diagrams corresponding to the vertex function of Fig.~\ref{fig:vertex}}
\label{fig:real}
\end{figure}
Answering this question requires a computation of the
vertex function, starting at one loop. Rather than the vertex function
itself, however, it is more instructive to look at its real-emission
counterpart. Indeed, it is the imaginary part of the photon propagator
in the diagram of Fig.~\ref{fig:vertex} 
which leads to the double-logarithmic behavior
Eqs.~(\ref{offshellone}-\ref{onshellall})\cite{Jackiw:1968zz,Berestetsky:1982aq}. This
imaginary part  can be extracted using the standard cutting rule
\begin{equation}
  {\rm Disc}\frac{1}{k^2+i\epsilon}=-2\pi i \delta(k^2)
  \Theta(k^0).\label{cutkosky}
\end{equation}
This transforms the vertex function into the interference of  two
real emission diagrams (see Fig.~\ref{fig:real}). Of course, it  is
is this one-to-one correspondence of virtual and real emission
contributions which guarantees the cancellation of infrared
singularities for sufficiently inclusive physical
observables.\footnote{See Chapter~13 of Weinberg's treatise\cite{Weinberg:1995mt} for a
  modern discussion.}

I will first, present in Sect.~\ref{vertex} a direct computation of these real-emission
contributions, viewed as contributions to the decay amplitude of a virtual photon 
into a fermion-antifermion pair, in the rest frame of the virtual
photon. This computation reproduces the result of
Eqs.~(\ref{offshellone}-\ref{onshellone}), and it gives a first hint
on its origin. For a complete clarification, however, it is useful to
look at
the problem in a different frame: namely, by viewing the diagrams of
Fig.~\ref{fig:real} as contributions to Drell-Yan-like production of
a virtual photon in a fermion-antifermion collision, in the
center-of-mass frame of the colliding fermions. In Sect.~\ref{IRC} I
will show that the
origin of the difference between on-shell and off-shell can be traced
to a different
interplay of soft and collinear singularities in either
case. Exponentiation then ensues from the factorized structure of
phase-space, which can be proved using an argument\cite{Muselli:2017bad}
developed in order to combine soft and collinear resummation.

\section{Computing the vertex function\footnote{The computation
    presented in this section is based on unpublished notes by
    Paolo Nason\cite{nason}.} }\label{vertex}
Consider the decay of an off-shell photon with momentum $q$. Of
course, this means that the incoming fermion of Fig.~\ref{fig:real}
now is an outgoing antifermion, and the result in the kinematics of
Fig.~\ref{fig:real}  can be
recovered by crossing. However, because the double log behavior only
depends on the modulus of the momentum transfer, the computation can
be indifferently performed in either kinematics.
For simplicity we consider the case of massless fermions,
though it can be checked explicitly\cite{nason} that results are
unchanged with a finite fermion mass.

Double logs
Eqs.~(\ref{offshellone}-\ref{onshellone}) arise due to the soft region
of integration over the momentum $k$ of the emitted photon. The
amplitude for emission of a soft photon
is obtained multiplying the amplitude for the process without soft
photon by an eikonal factor\footnote{See Chapter~13 of
  Weinberg's treatise\cite{Weinberg:1995mt}.}, so that in this limit
the amplitude for the process of Fig.~\ref{fig:real} is given by
\begin{equation}\label{ampl}
  M= M_0 e \left( \frac{2 p_1^\mu}{(p_1+k)^2}- \frac{2 p_2^\mu}{(p_2+k)^2}\right),
\end{equation}
where $M_0$ is the amplitude without the extra photon. 
The square amplitude is then
\begin{equation}\label{amplsq}
  |M|^2= - |M_0|^2 e^2 \frac{8 p_1\cdot p_2}{(p_1+k)^2 (p_2+k)^2},
\end{equation}
and the desired real-emission amplitude is find integrating this over
the phase space of the emitted photon:
\begin{equation}\label{eq:phsp}
  d\Phi_k= \frac{k^2 dkd\cos\theta d\phi}{2E(2\pi)^3}=\frac{k
    dEd\cos\theta}{8\pi^2},
  \end{equation}
where $E$ and $k$ are respectively the energy and modulus of the
three-momentum of the emitted photon, and in the last step we have
used $\frac{dE}{k}=\frac{dk}{E}$, and integrated over the azimuth $\phi$.

The calculation is performed in the on-shell case by assuming a
small photon mass $\mu$ as a regulator, so $(p_i+k)^2=2p_i\cdot
k+\mu^2$, and in the off-shell case by assuming $p_i^2>0$ so
$(p_i+k)^2=2p_i\cdot k+p_i^2$.

\subsection{On-shell fermions}

In the rest frame of the decaying photon the square amplitude is
\begin{align}\label{amplsqona}
  |M|^2&= - |M_0|^2 e^2 \frac{4 s}{\left[ E \sqrt{s}(1-\beta
      \cos\theta)+\mu^2 \right] \left[ E \sqrt{s} (1+\beta
      \cos\theta)+\mu^2 \right]}\\ \label{amplsqonb}
  &=   - |M_0|^2 e^2   \frac{4}{E^2 \left[ (1-\beta
      \cos\theta)+\frac{\mu^2}{E\sqrt{s}} \right] \left[  (1+\beta
      \cos\theta)+\frac{\mu^2}{E\sqrt{s}}\right]},
\end{align}
where $E=k^0$ is the emitted photon's energy, $\sqrt
{s}=\sqrt{(p_1+p_2)^2}=2 p_i^0=\sqrt{|q^2|}$ is the energy of the
fermion-antifermion system, and
\begin{equation}\label{betadef}
\beta=\sqrt{1-\frac{\mu^2}{E^2}}=1-\frac{\mu^2}{2 E^2}\left(1+O(\mu^2/E^2)\right).
\end{equation}

Integrating over the emitted photon's phase space Eq.~(\ref{eq:phsp}) 
we get
\begin{align}\label{amplsqonint}
  &\int   d\Phi_k |M|^2\nonumber\\
  &\quad= - |M_0|^2 \frac{e^2}{2\pi^2}\int \frac{EdEd\cos\theta}{E^2 \left[(1-\cos\theta+\cos\theta\frac{\mu^2}{2E^2}+\frac{\mu^2}{E
      \sqrt{s}})\right]\left[(1+ \cos\theta -\cos\theta\frac{\mu^2}{2E^2}+\frac{\mu^2}{E
      \sqrt{s}})\right]}
\end{align}
This leads to logarithmic behavior either when $\theta\to 1$ or
$\theta\to-1$, corresponding to the region in which the emitted
photon is respectively collinear to $p_1$ or $p_2$. These two
collinear and anti-collinear contributions are the same, and we get
\begin{align}\label{amplsqonint1}
\int   d\Phi_k |M|^2=- 2|M_0|^2 \frac{ e^2}{4\pi^2}\int 
\frac{dE d\cos\theta}{E\left(1-\cos\theta \right)}+{\rm non\> log},
\end{align}
where we have for definiteness written the collinear contribution, while
introducing a factor of two in order to account for the anticollinear
one, and we have retained the leading term in an expansion in $\cos\theta$ about
$\cos\theta=1$, as well as  in an expansion of $\mu^2$
about $m^2=0$, so the second square bracket in the denominator of
Eq.~(\ref{amplsqonint}) just
reduces to a factor of two.

Performing the angular integral we immediately get
\begin{equation}\label{amplsqoninte}
\int  d\Phi_k |M|^2=-2 |M_0|^2
\frac{\alpha}{\pi}\int\frac{1}{2}\frac{d E^2}{E^2}\ln\left[ \left(\frac{\mu^2}{2E^2}+\frac{\mu^2}{E\sqrt{s}}\right)^{-1}\right],
\end{equation}
where we have introduced the fine-structure constant $\alpha=\frac{e^2}{4\pi}$.
The double integral comes from the infrared region of integration over
the energy $E$ of the emitted fermion, hence we can neglect the second
term in the argument of the log, and we get, keeping only double
logarithmic terms,
\begin{equation}\label{amplsqonintfin}
\int   d\Phi_k|M|^2=- |M_0|^2
\frac{\alpha}{2\pi}\ln^2 \frac{|q|^2}{\mu^2},
\end{equation}
where the upper limit of integration over energy is of course just
$\sqrt s/2$; indeed, the argument of the log is fixed by dimensional
analysis.

The real emission contribution should be compared to the square of
the virtual one, which leads to an extra factor of two in the real
emission case, so this result exactly matches Roman's
result\cite{Jackiw:1968zz} for the on-shell vertex
function Eq.~(\ref{onshellone}).

\subsection{Off-shell fermions}
If the fermions are off-shell, it is now the fermion virtuality which
regulates the collinear singularity, so that no massive photon
regulator is needed. The square amplitude is then
\begin{align}\label{amplsqoffa}
  |M|^2&= - |M_0|^2 e^2 \frac{4 s}{\left[ E \sqrt{s}(1-\beta
      \cos\theta)+p_1^2 \right] \left[ E \sqrt{s} (1+\beta
      \cos\theta)+p_2^2 \right]}\left[1+O(p_i^2/s)\right]
\end{align}
where now the equalities $2 p^0_i\approx \sqrt{|q^2|}\approx\sqrt
{s}$ all hold up to terms of order $p_i^2/s$, and 
\begin{equation}\label{betaidef}
\beta_i=\sqrt{1-\frac{p_i^2}{(p_i^0)^2}}=1-\frac{2 p_i^2}{s}\left[1 +O(p_i^2/s)\right].
\end{equation}

Integrating over the photon's phase space we have again a pair of
collinear and anticollinear singularities:
\begin{align}\label{amplsqoffintsud}
  &\int  d\Phi_k|M|^2\nonumber\\
  &\quad= - |M_0|^2 \frac{e^2}{2\pi^2}\int  \frac{E dEd\cos\theta}
{\left[ E (1-
      \cos\theta)+ \cos\theta E \frac{2 p_1^2}{s}  +
      \frac{p_1^2}{\sqrt{s}} \right]
\left[ E (1+
      \cos\theta)- \cos\theta E \frac{2 p_2^2}{s}  +
      \frac{p_2^2}{\sqrt{s}} \right]}.
\end{align}
However, there are two
differences: the form of the collinear cutoff, which now depends on
the virtuality, $p_i^2$ rather than  the photon mass $\mu^2$   and
also, the form of the
energy denominator --- the second square bracket in the denominator of
Eq.~(\ref{amplsqoffintsud}) --- which is now also cut off. Indeed, focusing as
before on the collinear contribution (with a factor of two accounting
for the anti-collinear one) we get
\begin{align}\label{amplsqoffint}
\int   d\Phi_k|M|^2=- 2|M_0|^2 \frac{
  e^2}{4\pi^2}\int 
\frac{E dE d\cos\theta}
{E \left[ (1-
      \cos\theta)+  \frac{2 p_1^2}{s}  +
      \frac{p_1^2}{E\sqrt{s}} \right]
  \left[  E +  \frac{2 p_2^2}{2\sqrt{s}} \right]}+{\rm non\> log},
\end{align}
where again we have kept the leading terms as $\cos\theta\to 1$.

Performing the angular integral we
now get 
\begin{equation}\label{amplsqoffinte}
\int   d\Phi_k|M|^2=-2 |M_0|^2
\frac{\alpha}{\pi}\int\frac{d E}{E +  \frac{p_2^2}{\sqrt{s} }}\ln\left[ \left( \frac{2 p_1^2}{s}+  \frac{p_1^2}{E\sqrt{s}}\right)^{-1}\right],
\end{equation}
so 
it is apparent that the logarithmic integration over $E$ is cut off by
$\frac{p_2^2}{\sqrt{s}}$. 
It is now the first term in the argument of the log which is
subleading, and  performing the integral over the energy gives
\begin{equation}\label{amplsqoffintfin}
\int   d\Phi_k|M|^2=- |M_0|^2\frac{\alpha}{\pi}\ln\frac{s}{p_1^2}\ln\frac{s}{p_2^2}=- |M_0|^2\frac{\alpha}{\pi}\ln\frac{|q|^2}{p_1^2}\ln\frac{|q|^2}{p_2^2}
\end{equation}
up to single logarithmic terms. This is indeed twice as big as the
on-shell result Eq.~(\ref{amplsqonintfin}), and thus it exactly
matches Roman's result\cite{Jackiw:1968zz}.
\section{Infrared and collinear singularities}\label{IRC}
Having reproduced the result of
Eqs.~(\ref{offshellone}, \ref{onshellone}), and in particular the factor
two difference between on- and off-shell, we would now like to understand
the origin of this difference. Comparing Eqs.~(\ref{amplsqonint}-\ref{amplsqoffint}),
it is
clear that in both cases the double log stems from a collinear
and an infrared singularity, respectively coming from the integral over
the angle and the energy of the emitted photon. The difference resides
in the way the singularities are regulated by the photon mass, or by
the virtuality: however, the factor two appears somewhat haphazard, as
it looks like the reason why  Eqs.~(\ref{onshellone}) is twice as
large is that the integration variable is the energy $E$, rather than
$E^2$, with the remaining $\sqrt{s}$ dependence contained in the
cutoff.

However, a more transparent physical interpretation appears if we
consider the same computation, but in a different frame. Namely, we
view the amplitude of Fig.~\ref{fig:real} as the production of an
off-shell photon in the annihilation of a fermion-antifermion pair, in
the center-of-mass reference frame of the incoming fermions. The
physics is then similar to the familiar one of Drell-Yan production in
QCD (in which the fermions are quarks). 

\subsection{The Sudakov parametrization}\label{sec:sud}

It is then convenient to introduce a Sudakov-like parametrization of
the momentum $k$ of the emitted photon (which, in the QCD analogy,
would be an emitted gluon):
\begin{align}\label{sudakovlc}
  k&= (1-x) \frac{p_1+p_2}{2}+ y \frac{p_1-p_2}{2}+ k_{\rm T},\\\label{sudakov}
& = x_1 p_1+ x_2 p_2 + k_{\rm T} \end{align}
where  $k_{\rm T}\cdot p_1=k_{\rm T}\cdot p_2=0$ is a space-like transverse
momentum vector, such that $k_T^2=-|k_T|^2$, and of course
\begin{align}\label{x1def}
  x_1&=\frac{1}{2}\left[ (1-x) + y\right],\\\label{x2def}
  x_2&= \frac{1}{2}\left[(1-x) -y\right], 
\end{align}
so that either $(x_1,x_2)$ or $(x,y)$ can be
used according to convenience.

In the center-of-mass
frame of the incoming fermion-antifermion pair the energy of the
emitted photon is
\begin{equation}\label{emen}
  E=(1-x)\frac{\sqrt{s}}{2}
  \end{equation}
while its longitudinal momentum component
\begin{equation}\label{longp}
k_z\equiv  y \frac{p_1-p_2}{2}
\end{equation}
is entirely fixed by the
    on-shell condition
\begin{equation}\label{empz}    
  |k_z|=\sqrt{E^2-|k_{\rm T}|^2}= y\frac{\sqrt{|p_1-p_2|^2}}{2}
\end{equation}
where in the general off-shell case
$|p_1-p_2|^2=s-2(p_1^2+p_2^2)$. Of course in the off-shell case $k_z$
Eq.~(\ref{empz}) is the longitudinal momentum only up to terms
proportional to the difference of the two virtualities. Solving for
$y$ we get
\begin{align}\label{yres}
  y&=\pm\sqrt{(1-x)^2-\frac{4|k_{\rm T}|^2}{s}}\left(1+ O(p_i^2/s)\right).
\end{align}

The advantage of this choice of parametrization is seen by writing the
phase-space of the emitted photon, which now takes the form
\begin{align}\label{phspsud}
  d\Phi_k=\frac{|k_{\rm T}|d |k_{\rm T}|d\phi dk_z}{2 E (2\pi)^3}=\frac{ d |k_{\rm T}|^2
    dE}{4 |k_z| (4\pi^2)},
\end{align}
instead of the previous Eq.~(\ref{eq:phsp}).
Using Eqs.~(\ref{emen},\ref{empz}) we get
\begin{align}\label{phspx}
 d\Phi_k=\frac{1}{4 (4\pi^2)}\frac{ dx d |k_{\rm T}|^2}{ \sqrt{(1-x)^2-\frac{4
       |k_{\rm T}|^2}{s}}}\left(1+O(p_i^2/s)\right).
\end{align}
This last form exposes the phase-space
origin\cite{Forte:2002ni,Muselli:2017bad} of the soft and collinear
singularity: 
if the squared amplitude behaves as
$|M|^2\twiddles{|k_{\rm T}|}\frac{1}{|k_{\rm T}|}$, the $k_{\rm T}$ integration is
logarithmic; but then as $|k_{\rm T}|\to0$ the square root factor in the
denominator reduces to $1-x$ and the $x$ integration also becomes
logarithmic, in the $x\to1$ limit in which the energy of the emitted
photon Eq.~(\ref{emen}) vanishes.

This can be exposed by
rewriting, in the limit as $|k_{\rm T}|^2\to 0$
\begin{align}\label{distribid}
d\Phi_k&=\frac{1}{4 (4\pi^2)} dx d |k_{\rm T}|^2\frac{ 1}{ \sqrt{(1-x)^2-\frac{4
      |k_{\rm T}|^2}{s}}}\nonumber\\
&=\frac{1}{4 (4\pi^2)} dx d |k_{\rm T}|^2\left[\frac{1}{(1-x)_+}-\frac{1}{2}\delta(1-x)
\ln\frac{4|k_{\rm T}|^2}{s}\right]+O(|k_{\rm T}|^2),
\end{align}
where we have introduced the standard plus distribution, implicitly
defined by the distributional identity
\begin{equation}\label{plusdef}
\int_0^1 dx \frac{1}{(1-x)_+} f(x)= \int_0^1 dx \frac{f(x)-f(1)}{1-x}. 
\end{equation}
Note that because $|k_{\rm T}|^2\le s/4$ the sign of the log in Eq.~(\ref{distribid})
is such that the
contribution proportional to the delta is always positive.
If, as mentioned, the squared amplitude behaves as
$|M|^2\sim\frac{1}{|k_{\rm T}|^2}$,  when integrating over the phase space
Eq.~(\ref{distribid}), the
$k_{\rm T}$ integration leads to a 
double log, which is now clearly seen to arise when both $|k_{\rm T}|^2$ but
also $x\to1$, because of the delta.

We now show this explicitly.
The amplitude has  the form of Eq.~(\ref{amplsq}), but with
$k\to-k$ because the fermions are in the final state, so
with the Sudakov parametrization Eq.~(\ref{sudakovlc}) 
\begin{equation}\label{amplsqsud}
  |M|^2= - |M_0|^2 e^2 \frac{8 p_1\cdot p_2}{(p_1^2-2k\cdot p_1)
    (p_2^2-2k\cdot p_2)},
\end{equation}
with
\begin{align}\label{eq:eikfact}
  k\cdot p_1&=\frac{1}{2}\left[(1-x)-y\right]p_1\cdot
  p_2+\frac{1}{2}\left[(1-x)+y\right] p_1^2=x_2 p_1\cdot p_2+x_1 p_1^2
  \nonumber\\
  k\cdot p_2&=\frac{1}{2}\left[(1-x)+y\right]\left(p_1\cdot p_2+p_2^2\right)
  =x_1 p_1\cdot p_2+x_2 p_1^2.
\end{align}
In the $|k_{\rm T}|\to0$ limit, using Eq.~(\ref{yres}) we get 
\begin{align}\label{x1smallkt}
  x_1&= \frac{1}{2}\left[(1-x)+y\right]=(1-x)+O(|k_{\rm T}|^2/s) \\\label{x2smallkt}
  x_2&=\frac{1}{2}\left[ (1-x) -y\right]= \frac{|k_{\rm T}|^2}{(1-x)s}(1+O(|k_{\rm T}|^2/s)),
\end{align}
where we have assumed for definiteness $y>0$, and the opposite sign
would simply amount to interchanging $x_1$ and $x_2$ (i.e. the
collinear ad anticollinear limits).
Equations~(\ref{x1smallkt}-\ref{x2smallkt}) show that $x_2\to0$
corresponds to the
collinear limit, while $x_1\to0$ to the soft limit, with the two
limits interchanged in the anticollinear case in which the negative
$y$ solution is chosen.  

\subsection{On-shell and off-shell}\label{sec:offon}

 For on-shell fermions, $2p_1\cdot p_2=s$ and $p_i^2=0$, so that
the matrix element is then given by
\begin{align}\label{amplsqon}
  |M|^2&= - 2 |M_0|^2 e^2  \frac{16}{s[(1-x)^2-y^2]}\nonumber\\
  &=- 2 |M_0|^2 e^2 \frac{4}{|k_{\rm T}|^2},
\end{align}
where we have used Eq.~(\ref{yres}) and we have provided a factor of 2
in order to account for the two solutions for $y$, which correspond
respectively to the collinear or anticollinear regions when $|k_{\rm T}|\to0$.

Hence, integrating over $x$ with the phase space Eq.~(\ref{distribid})
we get
\begin{align}\label{amponintsud}
\int   d\Phi_k|M|^2&= -|M_0|^2 \frac{e^2}{4 (4\pi^2)} \int dx d |k_{\rm T}|^2\left[\frac{1}{(1-x)_+}-\frac{1}{2}\delta(1-x)
\ln\frac{4|k_{\rm T}|^2}{s}\right]\frac{8}{|k_{\rm T}|^2}\\
&=- |M_0|^2\frac{\alpha}{2\pi} \ln^2\frac{s}{\mu^2},
\end{align}
where  the first equality holds up to non-logarithmic terms, and the
second equality, which  holds to double-logarithmic accuracy,
is found cutting  off the logarithmic integration over
$|k_{\rm T}|^2$ with an infrared regulator (photon mass) $\mu^2$; note that
the
sign follows from the fact that it is the lower limit of integration
which provides the $\mu^2$ dependence.

We thus get the same double-log result as
Eq.~(\ref{amplsqonintfin}). The advantage of this choice of frame is
that origin of the double log can be traced to the behavior of the phase space Eq.~(\ref{distribid})
in the simultaneous infrared $x\to1$ and collinear $|k_{\rm T}|\to 0$ limit.

Let us now turn to the off-shell case.
The denominator of the amplitude is now given by
\begin{align}\label{denoff}
  D&=(p_1^2-2k\cdot p_1)
  (p_2^2-2k\cdot p_2)\nonumber\\
  &=s[(1-x)^2-y^2]\left[\left(\frac{s}{2}+p_1^2\right)-\frac{p_1^2}{x_2}\right]\left[\left(\frac{s}{2}+p_2^2\right)-\frac{p_2^2}{x_1}\right]\left(1+O(p_i^2/s)\right). 
\end{align}
This immediately implies that when integrating with the phase space
Eq.~(\ref{distribid}) the term proportional to the delta does not
contribute: $x_1$ vanishes in the  $x\to1$ limit, so
$\lim_{x\to1} D=\infty$ 
because of the second factor in square brackets in Eq.~(\ref{denoff}).
Indeed, there is no longer an infrared singularity when $x\to 1$,
because the off-shellness regulates it.

It is then convenient to write the denominator as
\begin{align}\label{denoffa}
  D&=\left(x_2 \frac{s}{2}-p_1^2\right)\left(x_1 \frac{s}{2}-p_2^2\right)\left(1+O(p_i^2/s)\right)\nonumber\\
 &=\frac{1}{1-x} \left[ |k_{\rm T}|^2- p_1^2 (1-x)\right] s
  \left[(1-x)-\frac{p_2^2}{s}\right] \left(1+O(p_i^2/s)+O(|k_{\rm T}|^2/s)\right),
\end{align}
where in the second step we have used
Eqs.~(\ref{x1smallkt}-\ref{x2smallkt}), in the small $|k_{\rm T}|$ limit.

The integrated square amplitude is thus given by
\begin{align}\label{ampoffint}
\int   d\Phi_k|M|^2&= -|M_0|^2 \frac{e^2}{4 (4\pi^2)} \int dx d
|k_{\rm T}|^2\frac{1}{(1-x)_+}\frac{4s}{D}\\
&=-2 |M_0|^2\frac{\alpha}{\pi} \int dx d
|k_{\rm T}|^2 \frac{1}{\left[ |k_{\rm T}|^2- p_1^2 (1-x)\right] 
  \left[(1-x)-\frac{p_2^2}{s}\right]}+{\rm non\> log},,
\end{align}
where again we have provided a factor of $2$ in order to account for
the two (collinear and anticollinear) solutions for $y$. Note that the
plus prescription in the first line of Eq.~(\ref{ampoffint}) has no
effect because the integrand vanishes at $x=1$, as it is clear from
Eq.~(\ref{denoffa}). 

The
integral over $|k_{\rm T}|^2$ in Eq.~(\ref{ampoffint})
is logarithmic about $ |k_{\rm T}|^2\sim p_1^2(1-x)$,
where the first factor in square brackets in
the denominator $D$ Eq.~(\ref{denoffa}) vanishes, thus leading to
\begin{equation}\label{ampoffintsud}
\int   d\Phi_k|M|^2=-2 |M_0|^2\frac{\alpha}{\pi} \int dx
  \ln\left(\frac{s}{2p_1^2(1-x)}\right)
  \frac{1}{(1-x)-\frac{p_2^2}{s}}.
  \end{equation}
The integral over $x$ has an infrared singularity
regulated by $p_2^2$ when $(1-x)\sim\frac{p_2^2}{s}$. The integral
over $x$ is thus again double-logarithmic, leading to
\begin{equation}\label{ampoffin}
\int   d\Phi_k|M|^2=-
|M_0|^2\frac{\alpha}{\pi}\ln\frac{s}{p_1^2}\ln\frac{s}{p_2^2}=-
|M_0|^2\frac{\alpha}{\pi}\ln\frac{|q|^2}{p_1^2}\ln\frac{|q|^2}{p_2^2}
\end{equation}
as in Eq.~(\ref{amplsqoffintfin}).

It is now clear that the factor two difference between the on-shell
result Eq.~(\ref{amponintsud}) and the off-shell result
Eq.~(\ref{ampoffin}) reveals a different underlying physics. In the
on-shell case, the double log stems from the soft-collinear region,
corresponding to the last term in the expression Eq.~(\ref{distribid})
of the phase space.  This is a genuine double log, in that it is due
to the square-root factor in the phase space being singular when both
the transverse momentum
$|k_{\rm T}|\to0$ and the energy of the emitted photon   $E\to0$ 
[i.e. $x\to1$, recalling 
 Eq.~(\ref{emen})]. In the off-shell case instead the double log is
really coming from the interference of two logarithmic integration
regions when the two propagators go on shell, with the phase space now
playing no role. These two integration regions correspond to an
integral over energy (or $x$) Eqs.~(\ref{phspsud}-\ref{phspx}) and
transverse momentum $|k_{\rm T}|$, but they are now decoupled.

\subsection{Exponentiation}\label{exp}
The argument presented in this Section so far concerns only the
one-loop or single-emission contributions of
Figs.~\ref{fig:vertex}-\ref{fig:real}, so one may wonder whether they
apply to all orders, and if so why.
Clearly, multiple eikonal emission does exponentiate, as textbook
arguments show, but the nontrivial question is what happens to the
phase space structure. However, it was recently\cite{Muselli:2017bad}
shown that the phase space for $n$-gluon (and thus also photon) emission  in the
small $|k_{\rm T}|$ has a factorized form which reproduces iteratively the
structure Eq.~(\ref{distribid}).

Specifically, the momenta of the emitted photons can be parametrized as
\begin{equation}
\label{eq:Sudakovpar}
k_i=\alpha_i\frac{p_1+p_2}{2}+y_i\frac{p_1-p_2}{2}+k_{\rm T}^i,
\end{equation}
so that of course
\begin{equation}
y_i=\pm\sqrt{\alpha_i^2-\frac{4|k_{t}^i|^2}{\hat s}}.
\label{alphabeta}
\end{equation}
Introducing
new variables $z_i$ through
\begin{equation}
\alpha_1=1-z_1;\qquad \alpha_i=z_1\ldots z_{i-1}(1-z_i),\;i\ge 2
\end{equation}
it can  then be shown\cite{Muselli:2017bad} that
\begin{equation}
\prod_{i=1}^n\frac{d\alpha_i}{\sqrt{\alpha_i^2-4\frac{|k_{\rm T}|^2}{s}}}
=\prod_{i=1}^n\frac{dz_i}{\sqrt{(1-z_i)^2-
\frac{4|k_{\rm T}|^2}{z_1^2\ldots z_{i-1}^2s}}},
\end{equation}
and that the phase space can be written as
\begin{align}
\label{eq:phasespacesmallpt4}
d\Phi_{n+1}\left(p_1,p_2;q,k_1,\dots,k_n\right)
&=\frac{8\pi^{3}}{\left[4(2\pi)^{2}\right]^{n+1}}
\frac{dq_t^2}{s} 
\int db^2\,J_0\left(b|q_t|\right)
\nonumber\\
&J_0\left(b|k_{\rm T}^1|\right)
    d|k_{\rm T}^1|^2dz_1
\left[\frac{1}{(1-z_1)_+}-\delta\left(1-z_1\right)\frac{1}{2}\ln\frac{|k_{\rm T}^1|^2}{s}\right]
\dots
\nonumber\\
&
J_0\left(b|k_{\rm T}^n|\right)
d|k_{\rm T}^n|^2 dz_n\,
\left[\frac{1}{(1-z_n)_+}-\delta\left(1-z_n\right)\frac{1}{2}\ln\frac{|k_{\rm T}^n|^2}{s}\right]
\nonumber\\
&\delta\left(\tau-z_1\dots z_n\right)+O\left(\frac{1}{b}\right),
\end{align}
where $\tau=\frac{|q^2|}{s}$.
The Fourier transform with respect to transverse momentum is necessary
in order to factorize the delta function which ensures transverse
momentum conservation, but it is clear that the structure of
Sect.~\ref{IRC} is then preserved and simply iterated, thereby leading
to exponentiation through arguments that are now textbook\cite{Peskin:1995ev} matter.

Clearly, once the problem is viewed in this way, the exponentiation is
seen to have the same origin, both on-shell and off-shell: even though
the different origin of the double log is manifested by the factor two
difference that we discussed.

As for the RG argument that Wilson asked Roman to
construct, it was eventually presented  thirty years
later\cite{Contopanagos:1996nh}, as a consequence of the factorization
of the integrated amplitude in terms of a factor which contains the
soft emissions, and the rest. The underlying physical reason
is\cite{Forte:2002ni} that in the soft limit the amplitude depends on
the variables $|k_{\rm T}|^2$, $x$ and $s$ only through the combination
$\frac{4 |k_{\rm T}|^2}{s(1-x)^2}$, essentially because $|k_{\rm T}^2|^{\rm
  max}=\frac{s(1-x)^2}{4}$ is the upper limit of the logarithmic transverse
momentum integration. The fact that the square amplitude only depends
on one variable then allows for RG improvement with
respect to it.

\section{Conclusion}\label{conc}
In summary, Roman's computation amounted to what in modern language
would be called the determination of the high-energy behavior of the
Drell-Yan process using reverse unitarity:~\cite{Anastasiou:2003gr}
the inclusive production of a gauge boson in fermion-antifermion
annihilation, computed using the cutting rule Eq.~(\ref{cutkosky}) to
express real radiation phase-space integrals in terms of loops.

What I have done here instead is to  compute the real emission
contributions directly. With this, I have   answered the
question that Roman had asked me: 
the factor two difference between the coefficient of the double
logarithm in  on-shell and off-shell  vertex functions (which then
 to all orders exponentiates) does have a simple physical
 interpretation. Namely,  in the on-shell case the double log is a
 soft-collinear log, coming from the behavior of phase-space in the
 region in which the energy and
 momentum of the emitted photon simultaneously go to zero; while
 in the off-shell case the double log is the interference of two
 independent logarithmic integrations, over the emitted photon energy
 and transverse momentum, coming from the region where  two
 propagator denominators vanish.
It is only in the on-shell case that there exists a genuinely infrared
and collinear region.

\bigskip

\section*{Added notes}

After publication on the {\tt arXiv} of the
first version of this paper, John Collins pointed out to me that the
factor of two disscussed in the present note is also discussed in his QCD
treatise~\cite{Collins:2011zzd}, and 
it was surely known in the early days of perturbative QCD:
indeed, it is mentioned (in passing)
in the introduction of the seminal paper by Mueller~\cite{Mueller:1979ih}
in which the Sudakov form factor is computed in QED on-shell, for
massive fermions (and photons),
 beyond the double
logarithmic approximation, to all logarithmic orders. This paper was
at the origin of subsequent generalizations to QCD by Collins
himself~\cite{Collins:1980ih}, which are at the basis of the
celebrated Collins-Soper-Sterman early results on QCD
factorization~\cite{Collins:1981uk,Collins:1981tt}.
Interestingly, Mueller's 
paper does altready introduce and exploit renormalization group
methods.

The discussion in the QCD book~\cite{Collins:2011zzd} presents a
computation of the loop diagram, similar to that performed by
Roman~\cite{Jackiw:1968zz} 
and presented in ``Landau''~\cite{Berestetsky:1982aq}; the origin of
the factor of two is  explained in terms of regions which
contribute to the loop integral in the on-shell vs. off-shell case
(see in particular Sect.~10.5.3 of the book~\cite{Collins:2011zzd}).
In this sense the computation presented here provides a complementary,
possibly more ``physical'' interpretation, to the extent that real
emission is physically more intuitive than virtual corrections.

Also, it was recently found that a curious difference of a
factor two in Sudakov form factors appears when
comparing initial and final state radiation~\cite{Forshaw:2019ver}. 

Interestingly, Mueller~\cite{Mueller:1979ih}  does not cite Roman's
result. We will leave it to the reader to draw a lesson from this
sequence of oblivions and re-discoveries --- including my own.

\medskip
{\it Note added in proof:} A few months after submission of this contribution, I was looking into
an old review paper that I should know quite well, by my late
mentor Guido Altarelli~\cite{Altarelli:1981ax}. I  realized that it
contains a section called ``The Sudakov form factor of partons'',
something that I had completely forgotten. I also realized that in
this section Eqs.~(\ref{offshellall}) and ~(\ref{onshellall}) are both to be found, accompanied by
the following sentence: ``It is well known that for an off shell quark
the exponent differs by a factor of 2''. No reference is given,
presumably because this fact is so well known.  
\smallskip

\section*{Acknowledgments}
I am grateful to Giancarlo Ferrera and Paolo Nason for several
discussions on the content of this paper, and also for a critical
reading of the manuscript. In particular, Giancarlo
pointed out to me the
discussion of the Sudakov form factor in Landau's treatise~\cite{Berestetsky:1982aq},  while the
computation presented in Sect.~\ref{sec:sudakov} is due to Paolo, whom I also
thank for providing me with detailed notes. I am very grateful to John
Collins for correspondence on the subject of this note and for calling
my attention to Refs.~\cite{Mueller:1979ih,Collins:1980ih} and
especially to the discussion in  his book~\cite{Collins:2011zzd}. I
also thank Giampiero Passarino for interesting comments and for
spotting some typos, Jeffrey Forshaw for pointing out his recent
work~\cite{Forshaw:2019ver}, and Phil Ratcliffe for spotting a typo in
an equation and for interesting comments on possible related work by
Landshoff and Stirling.\\
I acknowledge financial
support from  the European
Research Council under the European Union's Horizon
2020 research and innovation Programme (grant agreement n.~740006).
\bibliographystyle{UTPstyle}
\bibliography{sudakov2a}

\end{document}